\documentclass{elsarticle}

\usepackage{adjustbox}
\usepackage{array}
\usepackage{subfigure}
\usepackage{url}
\usepackage{pbox}
\let\labelindent\relax
\usepackage{enumitem}
\usepackage{xspace}
\usepackage[table,xcdraw,dvipsnames]{xcolor}
\usepackage{multirow}
\usepackage{longtable}
\usepackage{booktabs} 
\usepackage{multirow} 

\usepackage{amssymb}
\usepackage{epstopdf}
\usepackage{setspace}
\usepackage{lipsum}

\usepackage{todonotes}

\newenvironment{change}{\color{black}}{\color{black}}
\usepackage{xcolor}

\newcommand{\slowactions}{slow operations\xspace}

\begin{document}

\begin{frontmatter}

\title{In-The-Field Monitoring of Functional Calls: \\Is It Feasible?}

\author{Oscar Cornejo, Daniela Briola, Daniela Micucci, Leonardo Mariani}
\address{Department of Informatics, Systems and Communication\\
University of Milano - Bicocca, Milan, Italy}
\ead{{oscar.cornejo, daniela.briola, daniela.micucci, leonardo.mariani}@unimib.it}

\maketitle

\begin{abstract}


Collecting data about the sequences of function calls executed by an application while running in the field can be useful to a number of applications, including failure reproduction, profiling, and debugging. Unfortunately, collecting data from the field may introduce annoying slowdowns that negatively affect the quality of the user experience.

So far, the impact of monitoring has been mainly studied in terms of the overhead that it may introduce in the monitored applications, rather than considering if the introduced overhead can be really recognized by users. In this paper we take a different perspective studying to what extent collecting data about sequences of function calls may impact the quality of the user experience, producing recognizable effects. Interestingly we found that, depending on the nature of the executed operation and its execution context, users may tolerate a non-trivial overhead. This information can be potentially exploited to collect significant amount of data without annoying users.

\end{abstract}


\end{frontmatter}

\section{Introduction} \label{sec:introduction}

Behavioral information collected from the field can complement and complete the inherently partial knowledge about applications gained with in-house testing and analysis activities. For instance, observing applications that run in the field can produce data otherwise difficult to obtain, such as information about the behavior of the application when executed with actual production data and within various real execution environments. Indeed, collecting field data is common practice in the area of software experimentation~\cite{kohavi2013online,kevic2017characterizing,fagerholm2017right}, where controlled experiments are performed to evaluate how a change may impact the user experience.


    A diversity of data can be collected to study the behavior of software applications~\cite{Eclipse:website:2016,Windows:website:2016,Delgado:TaxonomyFaultMonitoring:TSE:2004,Jin:Sampling:SIGPLAN:2010,Liblit:BugIsolation:SIGPLAN:2003,Elbaum:Profiling:TSE:2005,Ohmann:OptimizedCoverage:ASE:2016,Pavlopoulou:ResidualCoverage:ICSE:1999,Jin:BugRedux:ICSE:2012,Clause:FieldFailuresDebugging:ICSE:2007}. In this paper, we focus on sequences of function calls, which is a specific but extremely common type of data recorded and used by analysis techniques. Studying the behavior of an application in terms of the function calls produced under different circumstances is in fact both common and useful. 
For example, sequences of function calls extracted from the field can be used to reproduce failures~\cite{Jin:BugRedux:ICSE:2012}, detect malicious behaviors~\cite{Gorji:MalwareDetection:ACMSE:2014}, debug applications~\cite{Murtaza:FaultLocalizationTheory:GTSE:2015}, profile software~\cite{Elbaum:Profiling:TSE:2005}, optimize applications~\cite{Zhao:Optimization:OOPSLA:2014}, and mine models~\cite{Mariani:Revolution:ISSRE:2012,Mariani:DynamicAnalysis:TSE:2011,7927993}.

Collecting information from the field is challenging since it slows down the application, and this may imply a negative effect on the quality of the user experience. If the slowdowns are frequent, the usability can be compromised up to the point users may stop using the application. It is thus extremely important to understand how the slowdowns introduced into an application can affect users. 

The impact of monitoring has been mainly studied in terms of its relative overhead, that is, by measuring how much the execution time of a given operation is increased due to the presence of the monitor. Although this is an important information, it does not reflect how and if this overhead can be perceived by the users of the application. For instance, increasing by 20\% the time that every menu item requires to open may introduce a small but annoying slowdown to operations that should be instantaneous from a user perspective. 
On the contrary, taking 20\% more time on the execution of a query might be acceptable for users, as long as the total time does not exceed their expectation. It is thus important to investigate the relation between the overhead introduced by monitoring techniques and the user experience, to understand how to seamlessly and feasibly collect data from the field.

In our initial study~\cite{NIER2017}, we discovered that a non-trivial overhead can be tolerated by users and that the overhead can be tolerated differently depending on the nature of the operation that is executed. This paper extends this initial study considering a larger number of operations exposed to overhead, new experiments to study how the availability of the computational resources may affect overhead, a study based on human subjects, and additional analyses of the empirical data. The results show that function calls can be frequently collected without impacting on the user experience, regardless of the availability of the computational resources, but specific operations may require ad-hoc support to be monitored without affecting users. These evidences can be exploited to design better monitor and analysis procedures running in the field.

This paper is organized as follows. Section~\ref{sec:experiment} describes our experimental setup. Sections~\ref{sec:results1} and~\ref{sec:results2} report the results obtained when studying the impact of the overhead on the users with good and poor availability of computational resources, respectively. Section~\ref{sec:empiricalstudy} describes the results obtained with our study involving human subjects. Section~\ref{sec:threats} discusses threats to validity. Section~\ref{sec:findings} summarizes our findings. Section~\ref{sec:related} discusses related work. Section~\ref{sec:conclusions} provides final remarks.

\section{Experiment Design} 
\label{sec:experiment}
This section describes the research questions that we have addressed and the design of the experiments to answer the research questions.

\subsection{Research Questions} 

The general objective of our study is understanding \emph{how collecting field data can affect the user experience}. We investigated this question in a specific, although common, scenario, that is, while recording the sequence of function calls executed by applications. 

We thus organized our study around three main research questions that investigate the impact of monitoring in different conditions. 

\smallskip
\noindent \textbf{RQ1 - How is the user experience affected by monitoring function calls?}  This research question analyzes the relation between the overhead produced by the monitoring activity and its impact on the user experience. RQ1 is further organized in three sub-research questions:

\smallskip
 \textbf{\emph{RQ1a - What is the overhead introduced by monitoring function calls?}}  RQ1a measures the overhead introduced in an application by the monitoring activity. 

\smallskip
 \textbf{\emph{RQ1b - What is the impact of monitoring function calls on the user experience?}} RQ1b studies if the overhead introduced with monitoring can be recognized by the user of the application.

\smallskip
 \textbf{\emph{RQ1c - What is the tolerance of the operations to the introduced overhead?}} RQ1c studies how different user operations tolerate overhead before producing slowdowns recognizable by users.

\smallskip
\textbf{\emph{RQ1d - Do failures change the overhead introduced by function calls monitoring?}} RQ1d studies if the overhead introduced by the monitor is different in the context of failures.

\medskip 

\noindent \textbf{RQ2 - What is the impact of monitoring function calls when the availability of computational resources is limited?} This research question investigates if, and how much, the overhead produced by collecting function calls changes with the availability of the computational resources. The study focuses on the availability of the two most relevant resources, CPU and memory, as captured by the following two sub-research questions:

\smallskip
 \textbf{\emph{RQ2a - What is the impact of CPU availability on the intrusiveness of monitoring?}}  RQ2a studies how the overhead introduced by monitoring function calls is affected by different levels of CPU utilization.
 
\smallskip
 \textbf{\emph{RQ2b - What is the impact of memory availability on the intrusiveness of monitoring?}} RQ2b studies how the overhead introduced by monitoring function calls is affected by different levels of memory utilization.

\medskip

Since we investigate RQ1 and RQ2 referring to the classification of the System Response Time as proposed by Seow~\cite{seow}, we consider the following research question to investigate the alignment between the user behavior and the adopted classification.

\medskip
	\noindent \textbf{RQ3 - How do \begin{change} expert computer\end{change} users react to the overhead produced by function calls monitoring, compared to the results obtained with RQ1 and RQ2?} This research question analyzes the alignment between the behavior of \begin{change}expert computer users recruited from our CS department \end{change} and the results reported in RQ1 and RQ2 with a study involving human subjects. 



\subsection{Experiment Design} 
This section presents the design of the experiment that we performed to answer to our research questions.  
To study the impact of monitoring we selected four widely used interactive applications: Notepad++ 6.9.2\footnote{\url{https://notepad-plus-plus.org}}, Paint.NET 4.0.12\footnote{\url{http://www.getpaint.net}}, VLC Media Player 2.2.4\footnote{\url{http://www.videolan.org}}, and Adobe Reader DC 2019\footnote{\url{http://get.adobe.com/reader}}. 


To collect sequences of function calls from these applications, we instrumented the applications using a probe that we implemented with the Intel Pin Binary Instrumentation Tool\footnote{\url{https://software.intel.com/en-us/articles/pin-a-dynamic-binary-}\linebreak\url{instrumentation-tool}}. 

Pin supports the instrumentation of compiled binaries, including shared libraries that are loaded at runtime, and optimizes performance  by automatically in-lining routines that have no control-flow changes~\cite{wallace2007superpin}. 
Our probe is a custom plug-in utility written in C++ 
that intercepts 
and logs every function call, included nested calls.

The probe uses a buffer of $50$MB to store data in memory before saving to file. We used this value based on the results we obtained in our preliminary experiment, where $50$MB resulted to produce the best compromise between CPU and memory consumption~\cite{NIER2017}.


To run each application, we have implemented a \textit{Sikulix}\footnote{\url{http://sikulix.com}} test case that can be automatically executed to run multiple functionalities of the monitored applications. The test cases simulate rich usage scenarios. 
For instance, Adobe Reader DC is executed by opening a PDF file, moving inside the document up and down several times, changing the view to full-screen, inserting comments in the text, searching for a specific word in the document, highlighting text, and closing the document. Notepad++ is executed by writing a Java program, opening different files, copying and pasting text in a document, counting the occurrences of a given word, marking the occurrences of a given word, and closing all the opened tabs. Paint.NET is executed by loading an image, resizing it, drawing several shapes and shaded shapes, rotating the image, applying different filters to the image (black and white, sepia), inverting the colors of the image, and closing it. VLC Media Player is executed by opening several video files, reproducing them, pausing them, adding and editing the subtitles of the video, playing videos from a given time, and creating and managing play-lists of different videos.

In the experiments, we first executed the test cases when the applications are not monitored and then with the monitor. To answer our RQs, we collected two main measures: the system response time overhead and the estimated impact on the user experience. We detail in the next section how we measured them. 

To accurately report the impact of the monitoring activity, we collected data about functions calls at the granularity of the individual operations performed in the tests.

An operation is a complete interaction between a user and an application: it starts with a user input (e.g., a user click) and ends with the application that has processed the input and is ready to accept the next input. For instance, an operation may start with a click on a menu and end with the menu being displayed. 

So, if a test case executes operations $o_1 \ldots o_n$, we collect functions calls and measure the overhead and its impact on the user experience for every operation $o_i$.  

Since our study targets interactive applications, we collect traces composed of user operations. 

It is possible to precisely distinguish the portion of the trace that corresponds to each operation by exploiting the knowledge of the name of the functions that implement the operations. This information is typically available if the organization that defines the monitoring strategy and the one that implements the application are the same. Otherwise, traces can still be split based on interactions with the GUI, but it implies a more sophisticated analysis of the collected traces.

 
To respond to RQ1, we only executed the test cases and the monitored applications, that is, no processes were running in addition to the basic operating system processes. To answer RQ2, we selectively saturated computational resources occupying $60\%$, $75\%$, and $90\%$ of both CPU and RAM. 

We performed linear sampling of the memory because we could not predict when there would be observable consequences. We considered saturation up to 90\% for both resources, since higher values would not allow us to satisfy the minimum requirements of Pin.

To saturate resources in a controlled way we used CPUStress 1.0.0.1\footnote{\url{https://blogs.msdn.microsoft.com/vijaysk/2012/10/26/tools-to-simulate-cpu-memory-disk-load/}} and HeavyLoad 3.4\footnote{\url{https://www.jam-software.com/heavyload/}}. %
To mitigate any effect due to non-determinism, we repeated each test 5 times and reported mean values. The overall study implied collecting and processing more than 10.000 samples about operations and their duration, all available at {\small \url{http://github.com/ocornejo/fieldmonitoringfeasibility}}.

\subsection{Measuring Overhead and Its Estimated Impact on the User Experience} 

Measuring the \emph{overhead} is straightforward, that is, we measure the difference in the duration of the same operations when executed with and without monitoring. Here it is important to discuss how we estimated the effect of the monitor on the user experience. In principle, assessing if a given overhead may or may not annoy users requires direct user involvement. However, user studies are expensive and can be hardly designed to cope with a volume of samples like the ones that we collected, which would require involving users in the evaluation of the duration of thousands of operations.

To estimate the impact of the overhead on users we thus exploited the results already available from the human-computer interaction domain, and we strengthen the collected evidence with a human study focusing on a restricted number of cases. In particular, we used the well-known and widely accepted classification proposed by Seow~\cite{seow} of the System Response Time (SRT, i.e., the time taken by an application to respond to a user request) that can be associated with each operation based on its nature. In this classification, operations are organized according to four categories, which have been derived from direct user engagement:

\begin{itemize}[leftmargin=*]
\item \emph{Instantaneous}: these are the most simple operations that can be performed on an application, such as entering inputs or navigating throughout menus. Users expect to receive a response by $100-200$ms at most.
	\item \emph{Immediate}: these are operations that are expected to generate acknowledgments or very simple outputs. Users expect to receive a response by $0.5-1$s at most.
	\item \emph{Continuous}: these are operations that are requested to produce results within a short time frame to not interrupt the dialog with the user. They are expected to produce a response in $2-5$s at most, depending on the complexity of the operation that is executed. We assume Simple Continuous operation to produce a response by $2-3.5$s and more Complex Continuous operations to produce a response by $3.5-5$s. 
	\item \emph{Captive}: these are operations requiring some relevant processing for which users will wait for results, but will also give up if a response is not produced within a certain time. These operations are expected to produce a response by $7-10$s.
\end{itemize}

In our study, we estimate the impact of the overhead on users by measuring the number of operations that change their category due to overhead. Intuitively, an operation that takes the time that is usually taken by an operation of higher category to complete its execution is an operation that does not satisfy the expectation of the user. We refer to these operations as the \emph{slow operations}. Measuring the number of operations that become slow due to overhead provides an estimate of how often users are likely to be annoyed while using a monitored application. We attribute categories to the operations performed by the tests based on their execution time when no overhead is introduced in the system and considering the lower limit of the execution time of each category. For instance, operations that take at most $100$ms are classified as Instantaneous, while operations that take more than $100$ms but less than $0.5$s are classified as Immediate.
If the execution time of an operation executed while the application is monitored exceeds the \begin{change}lower\end{change} limit of its category, the operation is considered to be slow. 

This strategy allows us to use the SRT classification as a continuous scale, using the lowest limit for \begin{change}both\end{change} the categorization of the operations and the identification of the \emph{slow operations}. \begin{change}We thus obtained a conservative measure of the slow operations, that is, the real number of slow operations reported by users are likely to be lower than the ones reported with this metric.\end{change}

\smallskip
We use the \emph{overhead} and the \emph{number of slow operations} as the main variables to answer our research questions. 

\section{RQ1 - How is the user experience affected by monitoring function calls?} 
\label{sec:results1}

This section reports the results obtained for each sub-research question RQ1a-RQ1d, and finally discusses the overall results obtained for RQ1.

Since the monitored applications are desktop applications, we executed all the experiments in a machine running Windows 7 Pro with a 3.47 GHz Intel Xeon X5690 processor and 4 GB of RAM.

\subsection{RQ1a - What is the overhead introduced by monitoring function calls?}\label{subsec:rq1a}

 \begin{figure}
 	\centering
 	\includegraphics[width=1\textwidth]{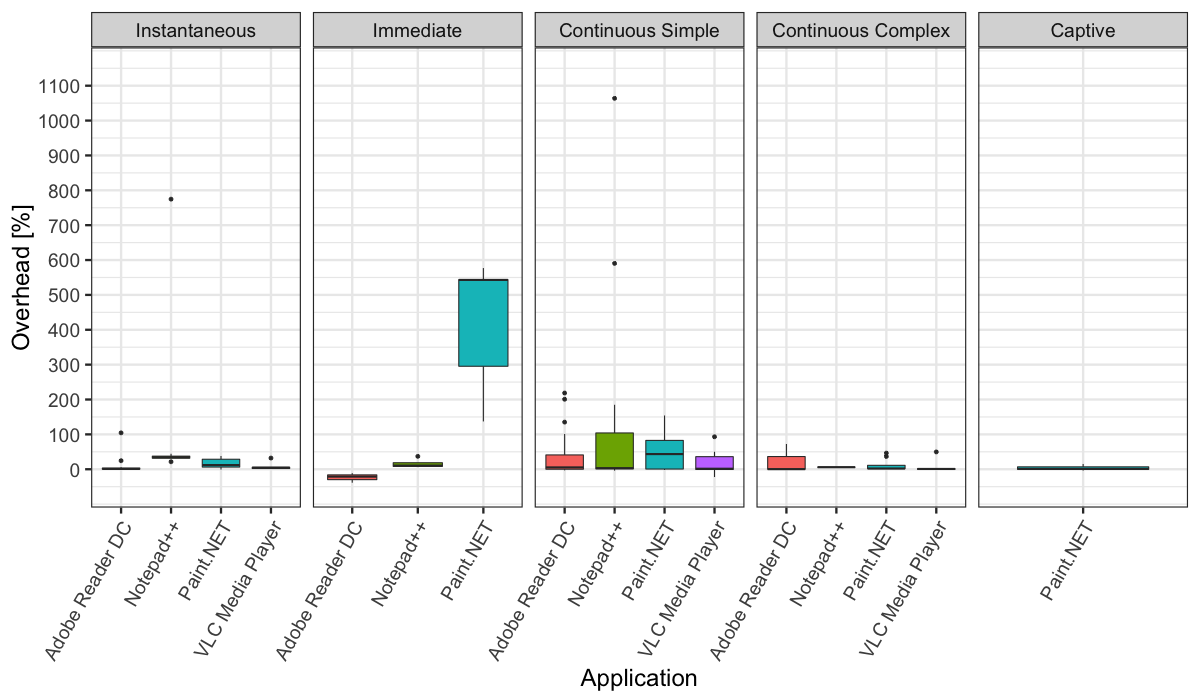} 
 	\caption{Overhead per category and application.}
 	\label{fig:overhead}
 \end{figure}

Figure~\ref{fig:overhead} shows the overhead that we observed for operations in each category and for each subject application. Note that not all types of operations occur in every application, for instance Captive operations are present in Paint.NET only.

The overhead profile per category is quite consistent. In the case of Instantaneous operations the overhead is always close to 0. This is probably due to the nature of Instantaneous operations that perform simple operations that imply the execution of a limited amount of logic and thus produce a limited number of function calls. A similar result can be observed for Immediate operations, where the overhead is small for Adobe Reader DC and Notepad++. Paint.NET represents an exception because its overhead is higher. The overhead profile is again quite consistent across operations in the Continuous Simple and Complex categories, with the overhead ranging between 0\% and 200\%. 

Although there are similarities for operations in the same category even if present in different applications, we can also observe that there are  exceptions. In fact, there are several outliers represented in the boxplot, with some of them showing very different overhead values compared to the rest of the samples. For example, we had two Continuous Simple operations in Notepad++ (selecting the Java highlighting and dismissing a save operation) with a high overhead (the two outliers) compared to the other operations, which experienced 100\% overhead at most.

\begin{figure}[!h]
\centering
\includegraphics[width=\textwidth]{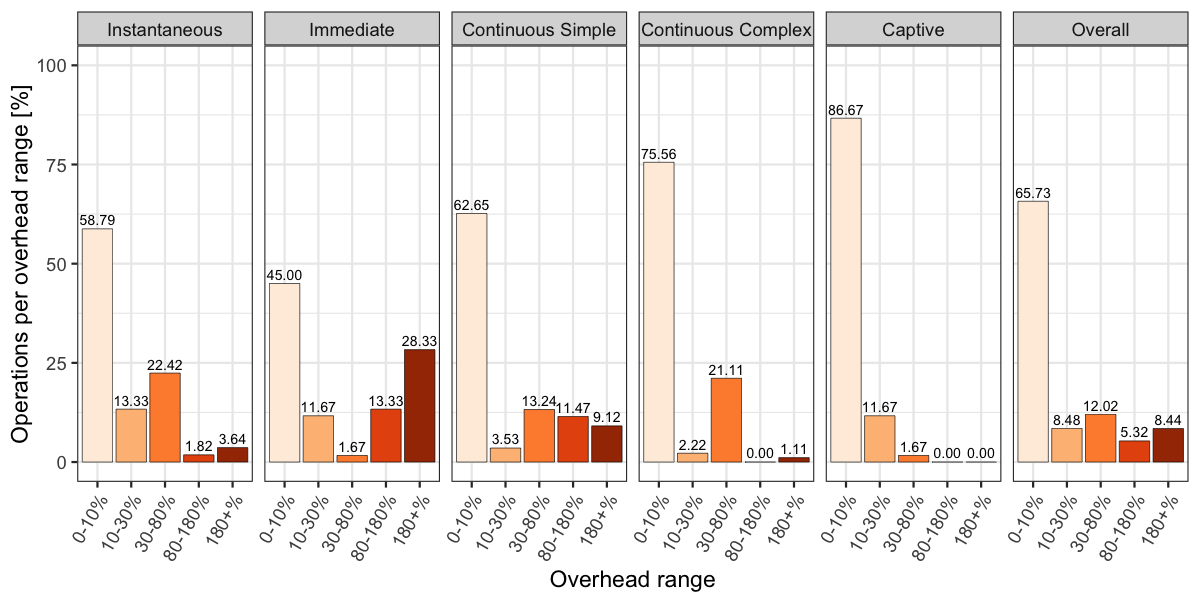}
\caption{Percentage of operations undergoing a specific overhead interval.}
\label{fig:OverheadVSOutOfCat}
\end{figure}

Figure~\ref{fig:OverheadVSOutOfCat} shows the percentage of operations in each category affected by overhead levels within specific ranges. Collecting function calls produces an overhead in the interval 0-10\% in the majority of the cases (65\% of the executed operations). In 8\% of the cases, operations are exposed to an overhead between 10\% and 30\%. In 12\% of the cases monitoring produced an overhead in the interval 30-80\%, and for less than 15\% of the operations the overhead is higher.

\smallskip

\emph{We can conclude that the observed behavior within operations of a same category is not significantly different, although specific operations may violate this pattern (Figure \ref{fig:overhead}). Moreover, collecting function calls exposes operations to an overhead that is lower than 10\% in the large majority of cases, and is seldom higher than 80\% (Figure \ref{fig:OverheadVSOutOfCat}).} Estimating if and how much this overhead can be intrusive with respect to user activity is studied with the next research question.

\begin{table}[h]
\resizebox{\textwidth}{!}{%
  \centering

\begin{tabular}{lcccccccc}

	\textbf{Adobe Reader DC} & & & & & & & & \\ 
  
    \toprule                                                       
	 \begin{tabular}{c} \textbf{Operation} \\ \textbf{category} \end{tabular} 
	 & \textbf{Total}
	& \textbf{Instantaneous}
	& \textbf{Immediate}
	& \begin{tabular}{c} \textbf{Continuous} \\ \textbf{Simple} \end{tabular} 
	& \begin{tabular}{c} \textbf{Continuous} \\ \textbf{Complex} \end{tabular} 
	& \textbf{Captive}
	& $>$ \textbf{Captive}	
	& \begin{tabular}[c]{@{}c@{}} \textbf{Slow}\\\textbf{Operations [\%]}\end{tabular} \\ 
	 \midrule
	Instantaneous	& 55	& \cellcolor{lightgray}50	& \textbf{5}	& 0	& 0	& 0	& 0	& \textbf{9\%} \\ \midrule
	Immediate       & 15                        & 0                         & \cellcolor{lightgray}15                       & 0                                & 0                                 & 0                         & 0                                                         & \textbf{0\%}                                                      \\   \midrule
Continuous Simple    & 90                        & 0                         & 0                        & \cellcolor{lightgray}69                               &  \textbf{16}                                &  \textbf{5}                         & 0                                                         &  \textbf{23\%}                                                     \\   \midrule
Continuous Complex   & 15                        & 0                         & 0                        & 0                                & \cellcolor{lightgray}13                                &  \textbf{1}                         &  \textbf{1}                                                         &  \textbf{13\%}                                                      \\   \midrule
Captive         & 0                         & 0                         & 0                        & 0                                & 0                                 & \cellcolor{lightgray}0                         & 0                                                         &  \textbf{0\%}                                                      \\  
      \bottomrule

\\ \\

	\textbf{Notepad++} & & & & & & & & \\ 
      
      \toprule                                                       
	 \begin{tabular}{c} \textbf{Operation} \\ \textbf{category} \end{tabular} 
	 & \textbf{Total}
	& \textbf{Instantaneous}
	& \textbf{Immediate}
	& \begin{tabular}{c} \textbf{Continuous} \\ \textbf{Simple} \end{tabular} 
	& \begin{tabular}{c} \textbf{Continuous} \\ \textbf{Complex} \end{tabular} 
	& \textbf{Captive}
	& $>$ \textbf{Captive}	
	& \begin{tabular}[c]{@{}c@{}} \textbf{Slow}\\\textbf{Operations [\%]}\end{tabular} \\ 
	 \midrule

Instantaneous   & 45    & \cellcolor{lightgray}40    & 0    &  \textbf{5}            & 0             & 0     & 0        &                                                 \textbf{11\%} \\ \midrule
Immediate       & 20    & 0     & \cellcolor{lightgray}19   &  \textbf{1}            & 0             & 0     & 0            &                                             \textbf{5\%} \\ \midrule
Continuous Simple    & 70    & 0     & 0    & \cellcolor{lightgray}48           & \textbf{6}             &  \textbf{11}    &  \textbf{5}     &                                                    \textbf{31\%} \\ \midrule
Continuous Complex   & 5     & 0     & 0    & 0            & \cellcolor{lightgray}5             & 0     & 0                             &                            \textbf{0\%} \\ \midrule
Captive         & 0     & 0     & 0    & 0            & 0             &  \cellcolor{lightgray}0     & 0                                               &          \textbf{0\%} \\
      \bottomrule

\\ \\

	\textbf{Paint.NET} & & & & & & & & \\ 
      
      \toprule
	 \begin{tabular}{c} \textbf{Operation} \\ \textbf{category} \end{tabular} 
	 & \textbf{Total}
	& \textbf{Instantaneous}
	& \textbf{Immediate}
	& \begin{tabular}{c} \textbf{Continuous} \\ \textbf{Simple} \end{tabular} 
	& \begin{tabular}{c} \textbf{Continuous} \\ \textbf{Complex} \end{tabular} 
	& \textbf{Captive}
	& $>$ \textbf{Captive}	
	& \begin{tabular}[c]{@{}c@{}} \textbf{Slow}\\\textbf{Operations [\%]}\end{tabular} \\ 
	 \midrule
Instantaneous   & 35    & \cellcolor{lightgray}35    & 0    & 0            & 0             & 0       & 0  &  \textbf{0\%} \\ \midrule
Immediate       & 25    & 0     & \cellcolor{lightgray}0    & \textbf{24}           & 0             & \textbf{1}       & 0     &  \textbf{100\%} \\ \midrule
Continuous Simple    & 55    & 0     & 0    & \cellcolor{lightgray}45           & \textbf{7}             & \textbf{3}       & 0     &  \textbf{18\%}\\ \midrule
Continuous Complex   & 40    & 0     & 0    & 0            & \cellcolor{lightgray}29            & \textbf{11}      & 0   &  \textbf{28\%} \\ \midrule
Captive         & 60    & 0     & 0    & 0            & 0             & \cellcolor{lightgray}57      & \textbf{3}  & \textbf{5\%}    \\                           
      \bottomrule
\\ \\
	\textbf{VLC Media Player} & & & & & & & & \\ 
      
      \toprule
	 \begin{tabular}{c} \textbf{Operation} \\ \textbf{category} \end{tabular} 
	 & \textbf{Total}
	& \textbf{Instantaneous}
	& \textbf{Immediate}
	& \begin{tabular}{c} \textbf{Continuous} \\ \textbf{Simple} \end{tabular} 
	& \begin{tabular}{c} \textbf{Continuous} \\ \textbf{Complex} \end{tabular} 
	& \textbf{Captive}
	& $>$ \textbf{Captive}	
	& \begin{tabular}[c]{@{}c@{}} \textbf{Slow}\\\textbf{Operations [\%]}\end{tabular} \\ 
	 \midrule
Instantaneous   & 30    &  \cellcolor{lightgray}30    & 0    & 0            & 0             & 0     & 0                                                         & \textbf{0\%}                                                      \\   \midrule
Immediate       & 0     & 0     &  \cellcolor{lightgray}0    & 0            & 0             & 0     & 0                                                         & \textbf{0\%}                                                      \\   \midrule
Cont. Simple    & 125   & 0     & 0    &  \cellcolor{lightgray}99           &  \textbf{26}            & 0     & 0                                                         & \textbf{21\%}                                                      \\   \midrule
Cont. Complex   & 30    & 0     & 0    & 0            &  \cellcolor{lightgray}24            &  \textbf{6}     & 0                                                         &\textbf{20\%}                                                      \\   \midrule
Captive         & 0     & 0     & 0    & 0            & 0             &  \cellcolor{lightgray}0     & 0                                                         & \textbf{0\%}                                                      \\   \bottomrule
\end{tabular}

}
\caption{Slow operations per application}
\label{table:slow_operations}
\end{table}

\subsection{RQ1b - What is the impact of monitoring function calls on the user experience?}\label{subsec:rq1b}

Table~\ref{table:slow_operations} reports the analytical results obtained for the operations recorded as slow in the four subject applications. For each application the table shows  the number of operations in each category that have been executed in the experiment and how the operation has been classified once affected by the overhead caused by function calls monitoring. The overhead is not recognizable by users if the category does not change with monitoring overhead. A perfect result implies having all 0s outside the values in the diagonal (highlighted with a grey background). When an operation changes its category, the table shows what the new category of the operation is. The column  \emph{$>$ Captive} shows the number of operations whose duration is longer than the maximum allowed for a Captive operation. The last column, \emph{Slow Operations $[\%]$}, specifies the percentage of slow operations across all the executions. 

\begin{figure}
	\centering
	\includegraphics[width=\textwidth]{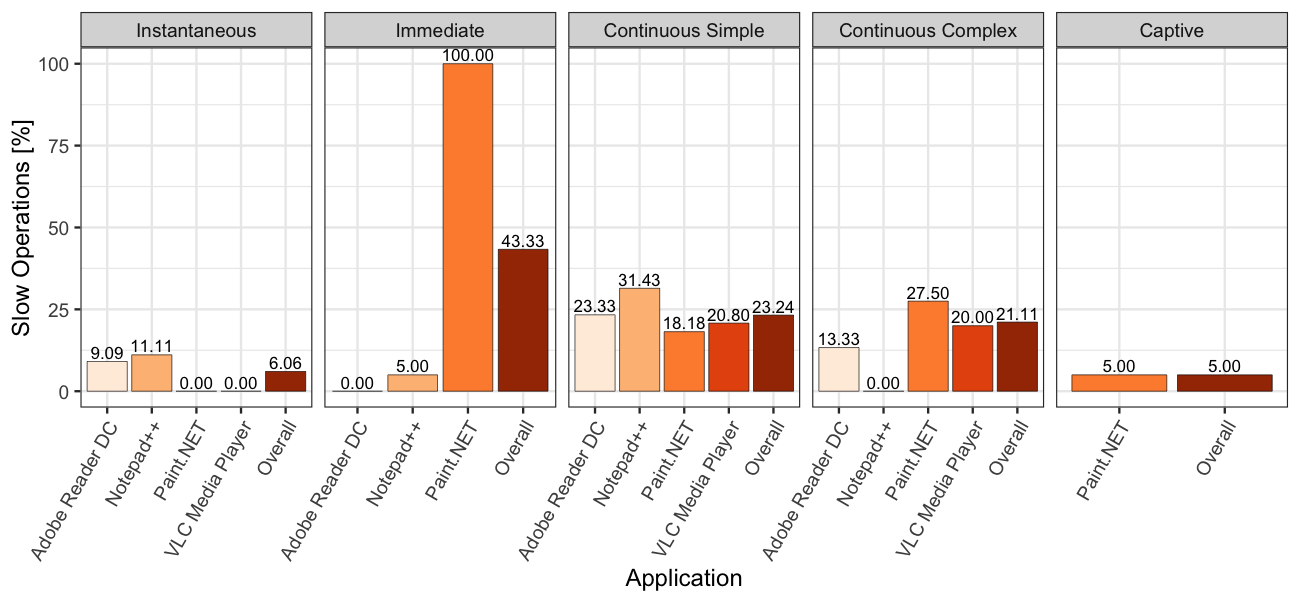} 
	\caption{Percentage of \slowactions with respect to the SRT Categories.}
	\vspace{-0.1cm}
	\label{fig:total}
\vspace{-0.2cm}
\end{figure}

Figure~\ref{fig:total} visually illustrates how slow operations distribute across operations categories. The last column in each category shows the percentage of slow operations for that category across all subject applications.

The empirical data suggests that Instantaneous operations seldom present a slowdown that affects the user experience: in fact only 6\% of the cases produced a recognizable slowdown. We obtained a similar result for Immediate operations with the exception of Paint.NET, where the slowdown has been significant for every Immediate operation that has been executed. This result is coherent with the exceptional overhead reported for Immediate operations in Paint.NET for RQ1a. This is likely caused by the nature of the Immediate operations in Paint.NET, which execute non trivial logic (e.g., the operation that closes an image) and are more expensive to monitor.

When the portion of logic of the application that is executed increases, the percentage of operations that become slow also increases, as observed for Continuous operations that in some cases become even slower than Captive operations (see Table~\ref{table:slow_operations}): for instance, the execution time of five Continuous Simple operations in Notepad++ exceeded the time expected for a Captive operation. The higher cost of monitoring Continuous operations is visible also in Figure~\ref{fig:total}, where more than 20\% of the Continuous operations (both Simple and Complex) have been significantly slowed down in average, compared to Instantaneous and Immediate operations where about 5\% of the operations have been slowed down, if we do not consider those from Paint.NET (which is a special case).

Extremely long tasks, such as Captive operations, seem to tolerate well the overhead caused by function calls monitoring. However, since they are present in one application only, it is hard to distill a more general lesson learnt. 

\medskip

\emph{We can conclude that the operations that are likely to be perceived as slowed down are quite limited in number ($<$20\% overall) and mostly concentrated in the Continuous operations. Moreover, applications that implement small pieces of logic that must be executed quickly, as Paint.NET does, might be particularly hard to monitor, in fact its Immediate operations have been all significantly slowed down when collecting function calls. }

\subsection{RQ1c - What is the tolerance of the operations to the introduced overhead?}\label{subsec:rq1c}

	\begin{figure}
	\centering
		\includegraphics[width=\textwidth]{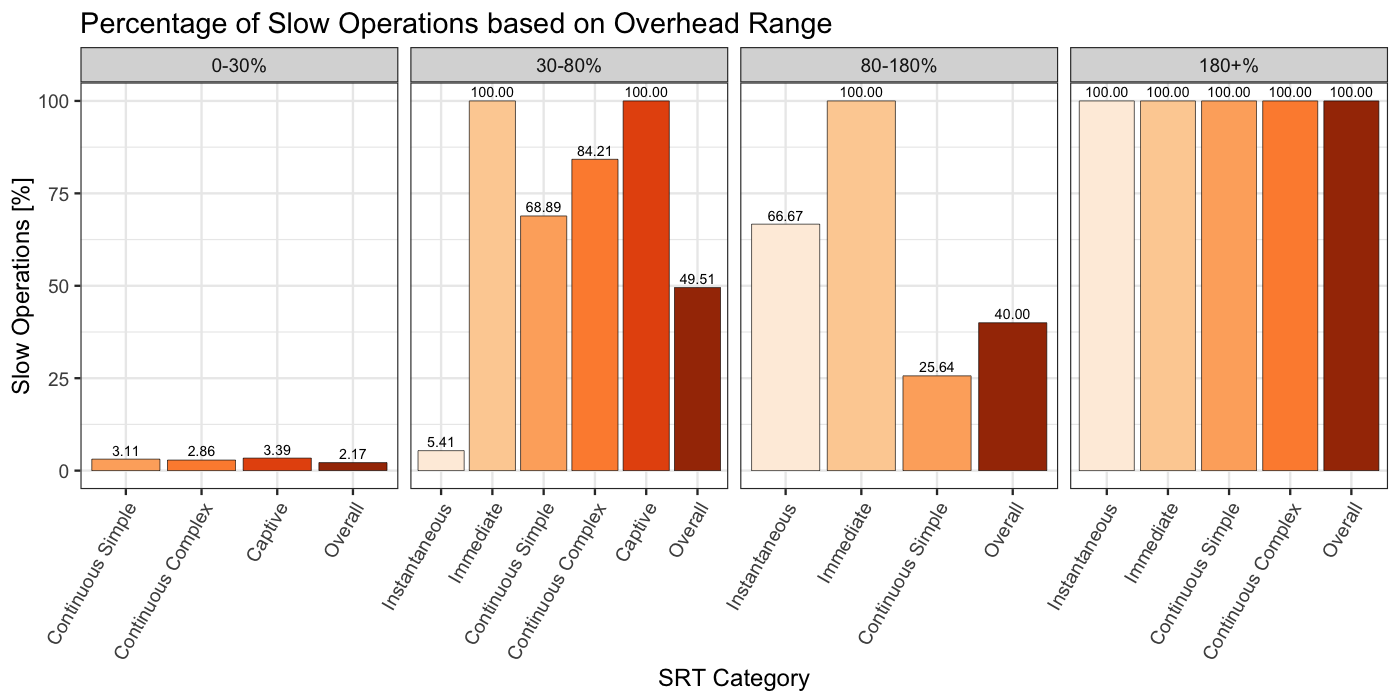} 
		\caption{Percentage of \slowactions for different overhead intervals.}
		\label{fig:catXrange}
	\end{figure}

Since we exposed operations in different categories to various overhead levels, this research question studies how often a certain overhead is the cause of operations resulting in a too slow response time. Figure~\ref{fig:catXrange} shows the percentage of operations, for all the categories, reported to be slow for overhead within a given range and for operations in all categories. 

\begin{change}
	In our previous study~\cite{NIER2017}, we identified 30\%, 80\%, and 180\% as  interesting overhead values that may produce different reactions by users, so we used these ranges in this study to analyze the collected data. 

We obtained a similar result with this experiment: 
an overhead level between 30\% and 80\% is hard to tolerate for operations in any category with the exception of Instantaneous operations, while overhead values higher than 80\% can be prohibitive. \end{change}


\medskip 

\emph{We can conclude that overhead levels up to 30\% are not harmful, but higher overhead levels must be introduced wisely with the exception of Instantaneous operations that seem to tolerate overhead slightly better than operations in the other categories.}


%
%


\subsection{RQ1d - Do failures change the overhead introduced by function calls monitoring?}\label{subsec:rq1d}

This research question investigates if monitoring functional calls may affect failures differently than regular executions. To compare the impact of monitoring when exactly the same operations terminate correctly or terminate with a failure, we inject faults into our subject applications. To this end, we configure PIN to modify the first instruction of a function if it is a MOV instruction with the AX register as a destination. The change consists of multiplying the destination address by a constant value. 

With this process, we achieved two applications failing abruptly (VLC Media Player and Paint.NET) and two applications presenting various misbehaviours (Adobe Reader DC and Notepad++). In the former case, the execution simply stopped prematurely, without producing any noticeable difference in terms of overhead. In the latter case, we obtained misbehaviors such as Adobe Reader DC failing to open files and Notepad++ failing to load graphical elements. We collected and analyzed the overhead values for   Adobe Reader DC and Notepad++.


\begin{figure}
\centering
	\includegraphics[width=\textwidth]{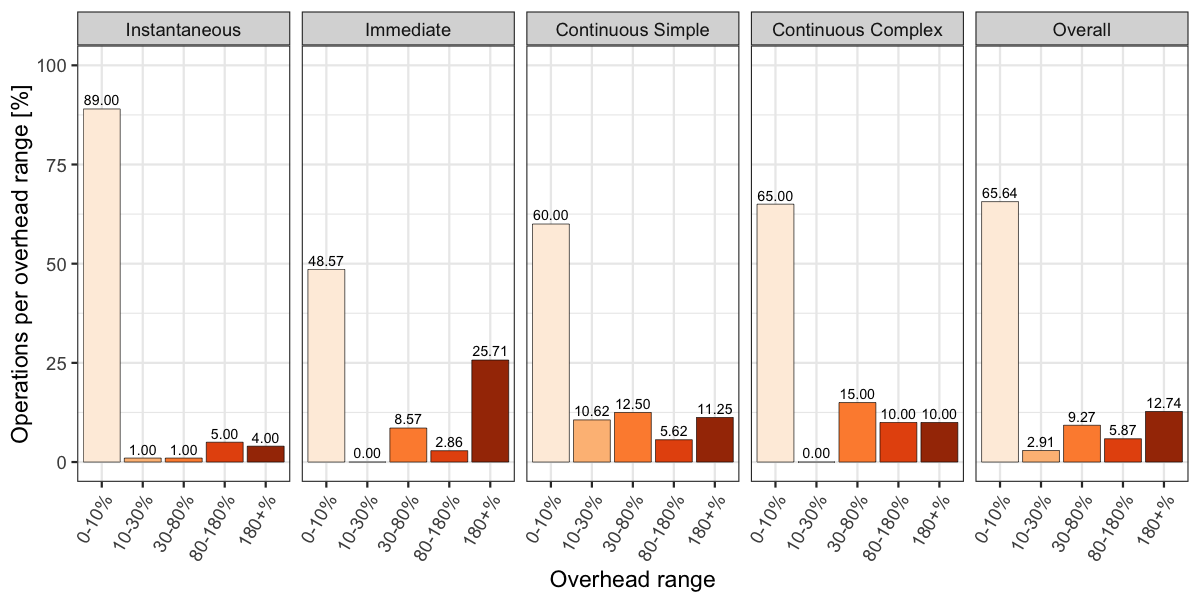} 
	\caption{Percentage of operations undergoing a specific overhead interval.}
	\label{fig:oh_catFaulty}
\end{figure}

Figure~\ref{fig:oh_catFaulty} shows the percentage of operations in each category affected by overhead levels within specific ranges. The result is very similar to the one presented in Figure~\ref{fig:OverheadVSOutOfCat} when the execution terminates correctly. In particular, collecting function calls during a failure produced an overhead in the interval 0-10\% in the majority of the cases for operation in any category (65\% of the operations that have been executed). We also observe that 2.91\% of operations produced an overhead in the range 10-30\%; 9.27\% of the operations produced an overhead in the interval 30-80\%; and for less than 20\% of operations the overhead was higher. 


%


\emph{In summary, failures do not change the cost of function calls monitoring, according to our observations.}

\subsection{Discussion}

Collecting function calls exposed the operations performed in the subject applications to various overhead levels: often below 10\% (65.73\% of the cases), and sometime to higher levels (8.48\%, 12.02\%, 13.77\%  of the cases in the ranges 10\%-30\%, 30\%-80\%, $>$80\%, respectively). 

We investigated if these overhead levels can be recognized by the users and we found that an overhead up to 30\% is well tolerated, while higher overhead levels can be tolerated for operations that usually execute fast. 


We do not consider the cost of aggregating and elaborating data on the client side, since techniques such as obfuscation~\cite{motiwalla2013developing,aggarwal2008privacy,ciriani2007microdata} and distributive monitoring~\cite{Bowring:MonitoringDeployedSoftware:PASTE:2002,Orso:GammaSystem:ISSTA:2002} usually consider this step as an offline process to be executed after the actual events have been collected, without impacting on the user experience.

These results suggest that monitoring activity, in particular collecting data about sequences of function calls, can be safely executed in many cases, but it must be controlled for those operations that execute an excessive amount of logic compared to their expected execution time. Monitoring techniques should be aware of these differences between operations and optimize their intrusiveness accordingly. 

Predicting and identifying operations expensive to monitor is an interesting challenge, which might be addressed with both static analysis and profiling techniques. Elaborating solutions in these directions is part of future work in this area.


\section{RQ2 - What is the impact of monitoring function calls when the availability of computational resources is limited?} 
\label{sec:results2}

In this section we study the impact of the monitoring activity when the computational resources cannot be completely allocated to the monitored applications but they are also allocated to other tasks. We first discuss the impact of CPU availability and then we discuss the impact of memory availability. 

Similarly to RQ1, we study the impact of collecting function calls by analyzing the overhead and studying the number of operations changing category when CPU and RAM are under stress.


\subsection{RQ2a - What is the impact of CPU availability on the intrusiveness of monitoring?}
\label{subsec:rq2a}


\begin{figure}[h]
\centering
\includegraphics[width=\textwidth]{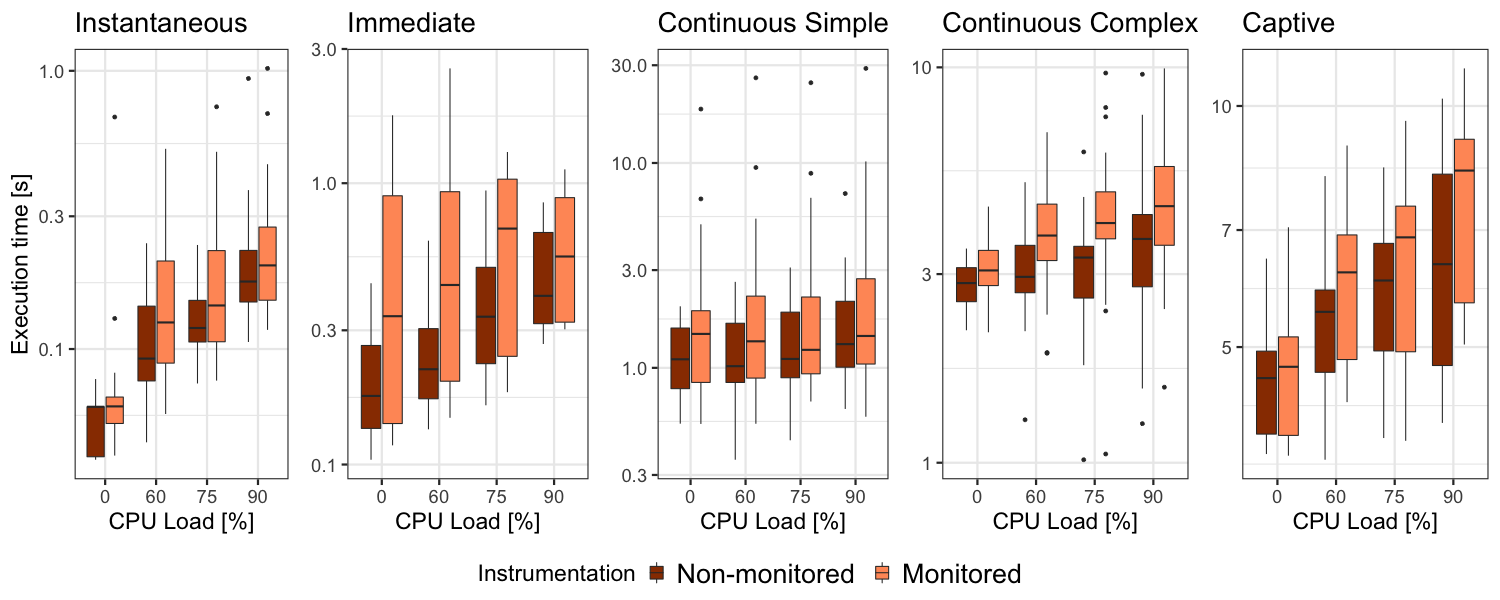}
\caption{Execution time for various CPU load levels per operation category.}
\label{fig:contextSRTCPU}
\end{figure}


\begin{change}
Figure~\ref{fig:contextSRTCPU} shows the system response time (presented in log scale) of the executed operations per operation category. We report timing information considering four CPU load levels: 0\%, 60\%, 75\%, and 90\%. 
\end{change}
The figure includes two types of boxplots: the orange boxplot corresponds to the execution time observed when monitoring is in place, while the brown boxplot corresponds to the execution time when no monitoring is in place.

\begin{table}[h]
\centering
\scriptsize
\begin{tabular}{*{4}{l}}
\toprule
\textbf{Treatment} & \textbf{Chi-square} & \textbf{{\it p}-value} & \textbf{df} \\
\midrule
Instantaneous      & 2.2107     & 0.5298  & 3  \\
Immediate          & 1.1327     & 0.7692  & 3  \\
Continuous Simple  & 3.3914     & 0.3351  & 3  \\
Continuous Complex & 4.4726     & 0.2147  & 3  \\
Captive            & 1.54       & 0.6731  & 3 \\
\bottomrule
\end{tabular}
\caption{Kruskal-Wallis test results per operation category.}
\label{table:cpuKruskal}
\end{table}

The trend is quite similar for all classes of operations with the exception of Immediate operations, which show decreasing values of the overhead for higher CPU load values. We conducted a Kruskal-Wallis test to check if the overhead introduced for a given CPU load and a given class of operations differs from the overhead for the same class of operations exposed to a different CPU load (significance expected for {\it p}-value $< 0.05$). The test revealed no significant differences (see Table~\ref{table:cpuKruskal}), suggesting that the impact of monitoring is not affected by a significant degree by the CPU load level, that is, an application is slowed down similarly by function calls monitoring regardless of the CPU availability.



\begin{figure}[h]
\centering
\includegraphics[width=\textwidth]{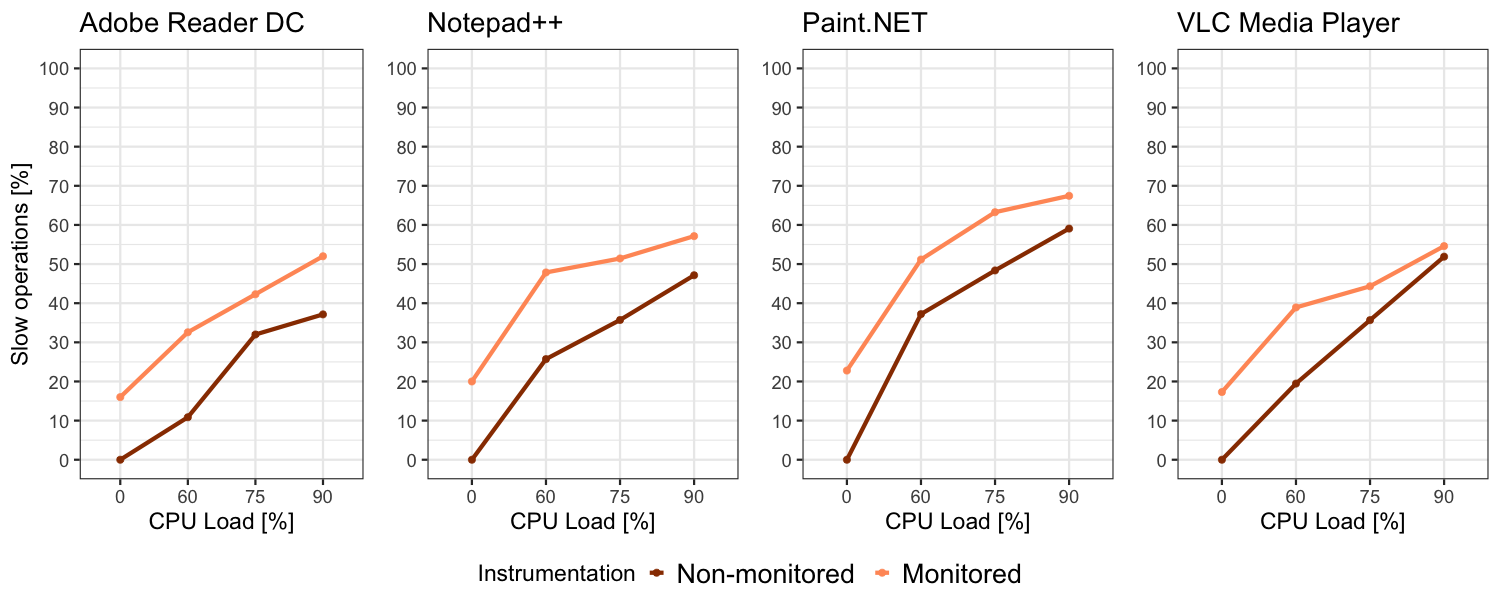}
\caption{Percentage of slow operations for various CPU load levels per application.}
\label{fig:contextAppsCPUslow}
\end{figure}

\begin{figure}[h]
\centering
\includegraphics[width=\textwidth]{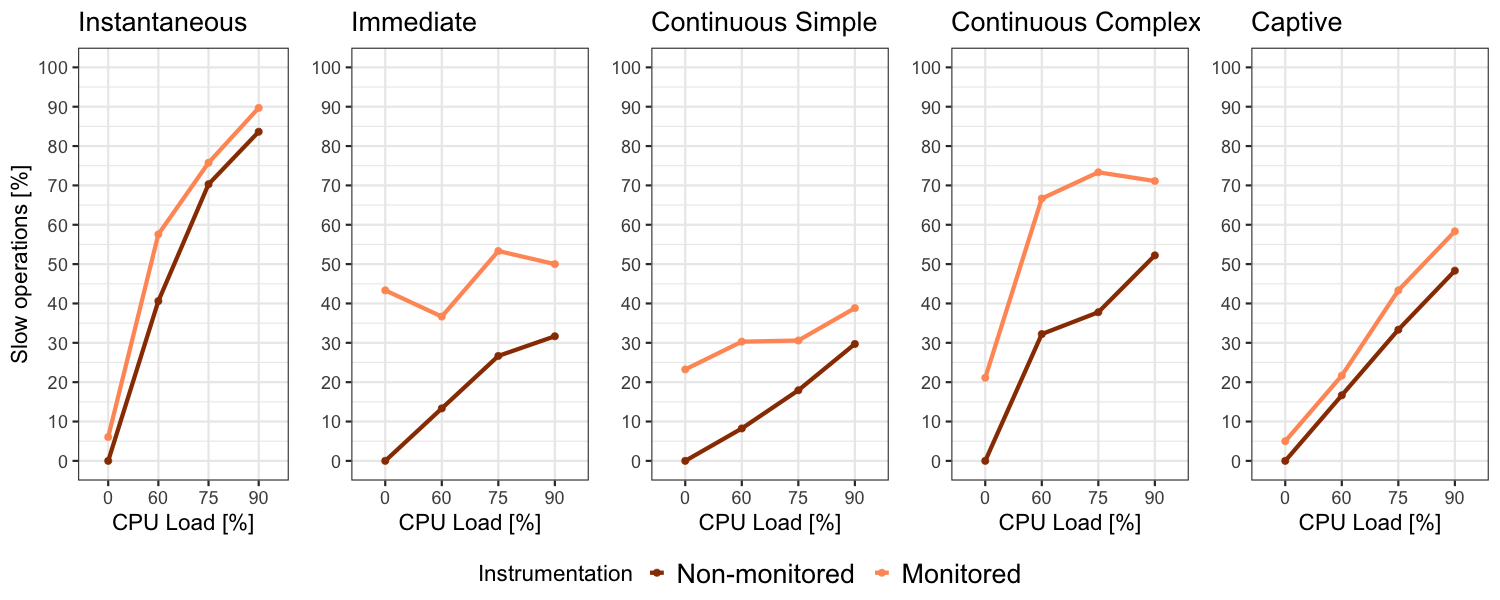}
\caption{Percentage of slow operations for various CPU load levels per operation category.}
\label{fig:contextSRTCPUslow}
\end{figure}

We also considered how monitoring affects the number of slow operations per application, shown in Figure~\ref{fig:contextAppsCPUslow}, and the number of slow operations per operation category, shown in Figure~\ref{fig:contextSRTCPUslow}. The usage of a loaded CPU already generates a number of slow operations for each application. Adding function calls monitoring further increases the number of operations that have been slowed down. We can however notice that the only addition of monitoring makes the user experience worse by a similar degree across CPU load levels, confirming that the CPU load level is not a significant factor when considering the impact of monitoring. To confirm this intuition we computed the linear regression of the number of slow operations for the instrumented and non-instrumented version of each application, and considered the difference between the angular coefficients of the computed lines. We further considered the percentage of operations with a different classification when the CPU saturates to 100\% (highest saturation possible) based on the computed trends. 
\begin{change}
Table~\ref{table:slopeAnalysisCPU} reports the results. For each application we indicate the difference between the angular coefficients (on the left) and the percentage of operations with a different categorization (on the right).
\end{change}

We can notice that the difference in the increase of the number of slow operations is between 2.66\% and 14.33\% of the operations, indicating a similar trend (i.e., slope) for the two cases (with and without monitoring). The small positive values of the difference between the coefficients indicates that, when a difference is observed (e.g., 14.33\% of the operations in Paint.NET), the saturation of the CPU increases the number of slow operations by a lower degree when monitoring is active.

\begin{table}[!b]
\scriptsize
\centering

\begin{tabular}{lcccc}
\toprule
 & \textbf{Adobe Reader DC} & \textbf{Notepad++} & \textbf{Paint.NET} & \textbf{VLC Media Player} \\
 \midrule
\textbf{CPU}                   & 0.047 -- 2.66\%         & 0.123 -- 8.78\%      & 0.308 -- 14.33\%      & 0.256 -- 13.85\%         \\
\bottomrule        
\end{tabular}

\caption{Trend analysis for CPU.}
\label{table:slopeAnalysisCPU}
\end{table}

The plot of the data per operation category, Figure~\ref{fig:contextSRTCPUslow}, reveals that Instantaneous operations behave better than the other operations in terms of their ability to tolerate monitoring, in fact the number of slow operations does not change significantly when monitoring is introduced in the system. Even if Captive operations behave similarly to Instantaneous operations, it is hard to generalize the result since they are present in one application only. On the other hand, Instantaneous operations are more sensitive to the load of the CPU, in fact, more than 80\% of the Instantaneous operations are slow when the CPU load reaches 90\%. 

\medskip

\emph{We can conclude that the CPU load level does not significantly affect the intrusiveness of function calls monitoring. In fact, the impact of the addition of monitoring tends to be the same regardless of CPU availability, and when a difference is observed, monitoring results to be slightly less intrusive with a higher saturation of the CPU.}

\subsection{RQ2b - What is the impact of memory availability on the intrusiveness of monitoring?}

\begin{table}[b]
\centering
\scriptsize
\begin{tabular}{*{5}{c}}
\toprule
    & \textbf{Adobe Reader DC} & \textbf{Notepad++} & \textbf{Paint.NET} & \textbf{VLC Media Player} \\
\midrule
Max RAM  & 356 MB    & 278 MB    & 520 MB    & 303 MB\\
\bottomrule
\end{tabular}
\caption{Maximum memory used during experimentation.}
\label{table:maxMemoryUsed}
\end{table}

To better discuss the results for RQ2b, we report the memory usage of each application as summarized in Table~\ref{table:maxMemoryUsed}: the maximum memory consumption observed during the execution of our tests for Adobe Reader DC is 353 MB of RAM, for Notepad++ is 278 MB of RAM, for Paint.NET is 520 MB of RAM, and for VLC Media Player is 303 MB of RAM.

\begin{figure}[h]
\centering
\includegraphics[width=\textwidth]{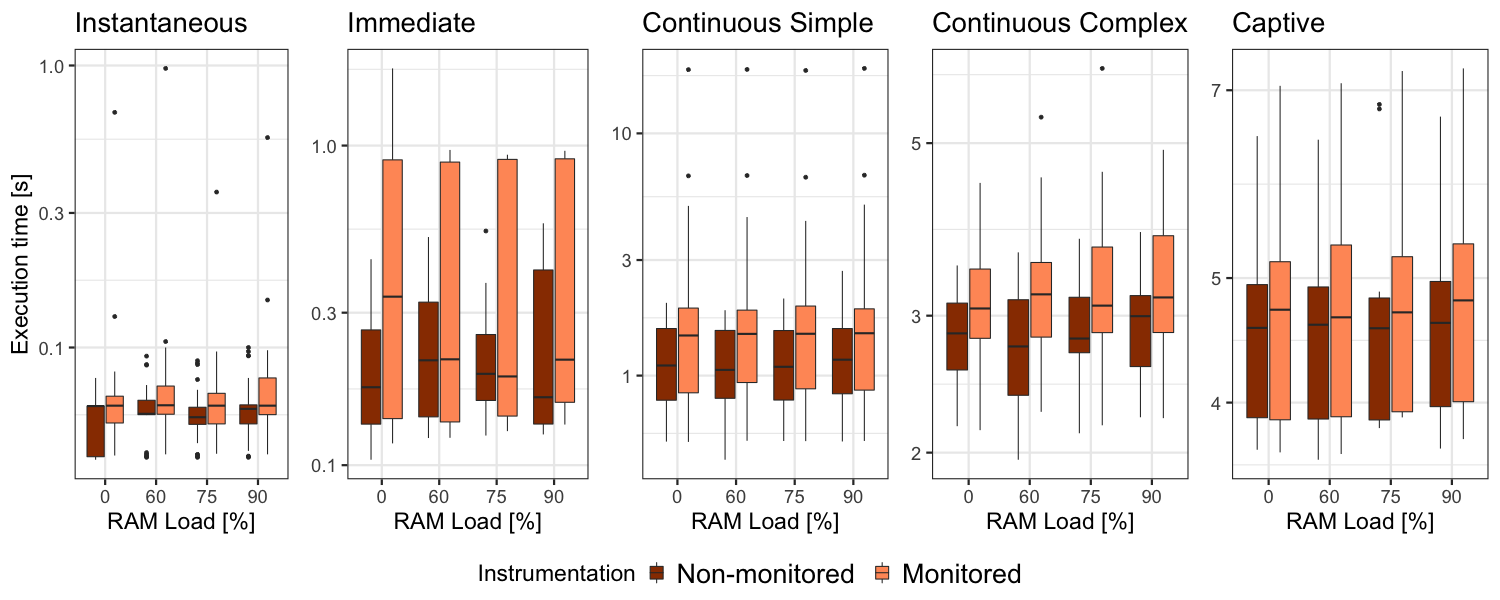}
\caption{Execution time for different RAM availability per operation category.}
\label{fig:contextSRTRAM}
\end{figure}

\begin{change}
	Figure~\ref{fig:contextSRTRAM} shows the overhead introduced in the system response time (presented in log scale) per operation category when varying the amount of occupied memory up to 90\%. 
\end{change}
The orange boxplot corresponds to the execution time observed when function calls are collected, while the brown boxplot corresponds to the execution time when no monitoring is in place. 

\begin{table}[h]

\centering
\scriptsize
\begin{tabular}{*{4}{l}}
\toprule
\textbf{Treatment} & \textbf{Chi-square} & \textbf{{\it p}-value} & \textbf{df} \\
\midrule
Instantaneous      & 3.5831     & 0.3101  & 3  \\
Immediate          & 3.5298     & 0.3169  & 3  \\
Continuous Simple  & 2.1604     & 0.5398  & 3  \\
Continuous Complex & 1.1438     & 0.7665  & 3  \\
Captive            & 0.1913     & 0.979   & 3 \\
\bottomrule
\end{tabular}
\caption{Kruskal-Wallis test results per operation category.}
\label{table:ramKruskal}

\end{table}

Similar to Section~\ref{subsec:rq2a} we check for statistical differences between groups using a Kruskal-Wallis test (see Table~\ref{table:ramKruskal}), obtaining no significant difference between different levels of RAM load.
Particularly, the results show a clearly negligible effect of the memory on the overhead, indeed the overhead is similar for different values of memory occupation.

\begin{figure}[h]
\centering
\includegraphics[width=\textwidth]{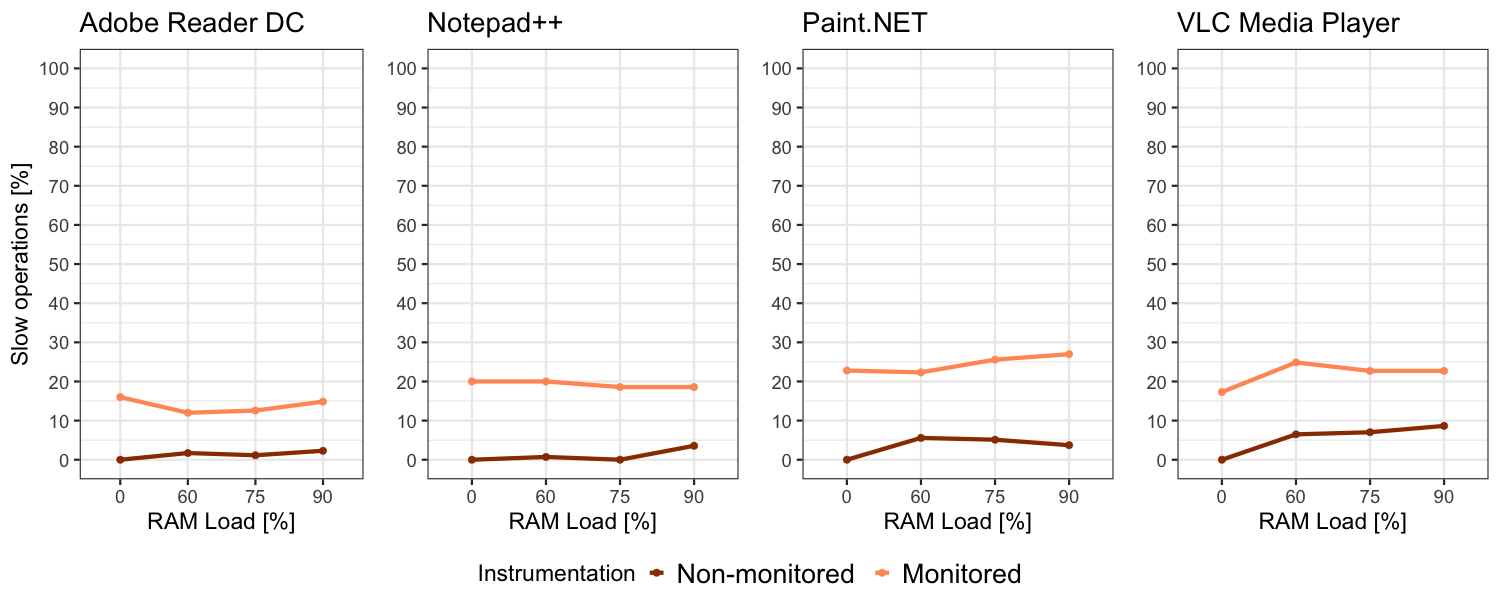}
\caption{Percentage of slow operations for various RAM load levels per application.}
\label{fig:contextAppsRAMslow}
\end{figure}

\begin{figure}[h]
\centering
\includegraphics[width=\textwidth]{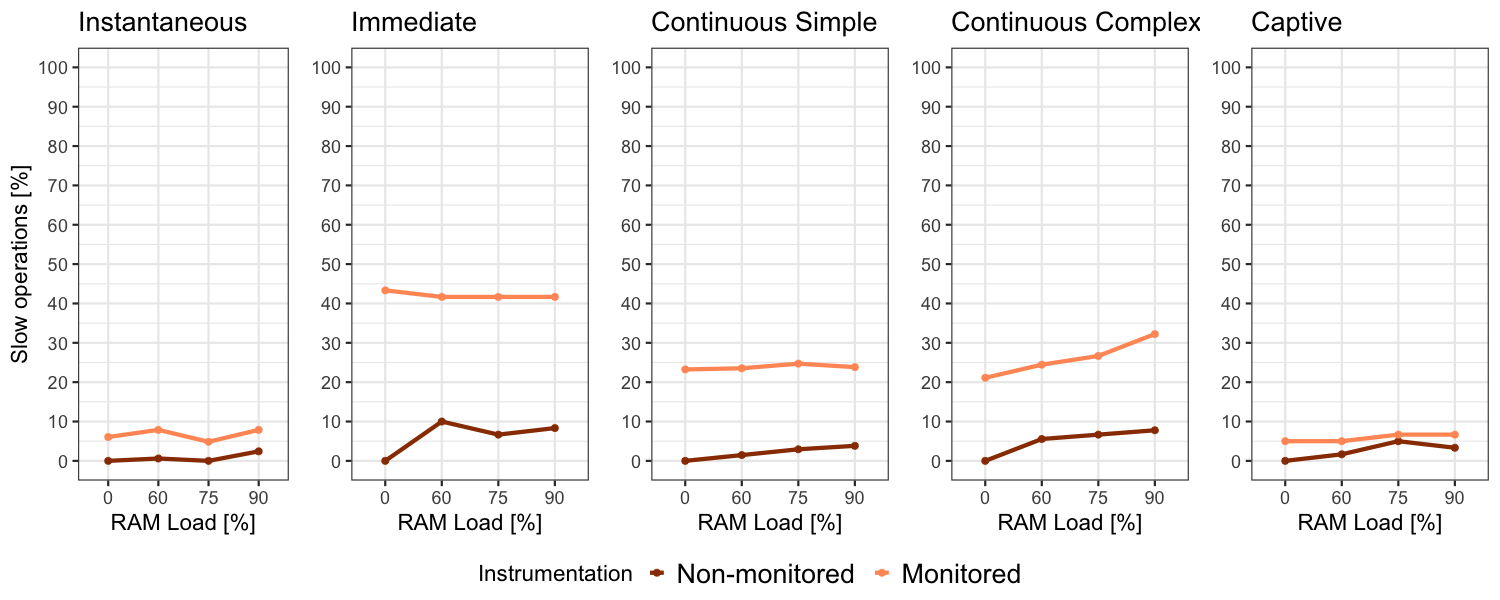}
\caption{Percentage of slow operations for various RAM load levels per operation category.}
\label{fig:contextSRTRAMslow}
\end{figure}

We also investigated how memory occupation impacts on the operations that become slow. Figure~\ref{fig:contextAppsRAMslow} shows the number of slow operations per application, while Figure~\ref{fig:contextSRTRAMslow} shows the number of slow operations per operation category. The behavior of the applications does not reveal any trend. To confirm this intuition we computed the linear regression of the number of slow operations for the instrumented and non-instrumented version of each app and considered the difference between the angular coefficients of the computed lines. We further considered the percentage of operations with a different classification when the memory saturates to 100\% (highest saturation possible) based on the computed trends. 
\begin{change}
Table~\ref{table:slopeAnalysisRAM} reports the results. For each application we indicate the difference between the angular coefficients (on the left) and the percentage of operations with a different categorization (on the right).
\end{change}
We can notice negligible difference in the coefficients and the number of slow operations, suggesting similar trend for the two cases.

The results per operation category confirm the same behavior we observed for CPU utilization: Instantaneous operations better tolerate low availability of the computational resources compared to operations in the other categories. Anyway, memory occupation does not produce relevant effects when analyzing the results per operation category, either.

\begin{table}[h]
\scriptsize
\centering

\begin{tabular}{lcccc}
\toprule
 & \textbf{Adobe Reader DC} & \textbf{Notepad++} & \textbf{Paint.NET} & \textbf{VLC Media Player} \\
 \midrule
\textbf{RAM}                   & 0.084 -- 4.77\%         & 0.059 -- 4.25\%      & 0.023 -- 1.08\%       & 0.054 -- 2.91\% \\
\bottomrule        
\end{tabular}

\caption{Memory trend analysis.}
\label{table:slopeAnalysisRAM}
\end{table}

\medskip

\emph{We can conclude that the memory load level does not affect the intrusiveness of function calls monitoring by a significant degree. In fact, the monitoring overhead tends to be the same regardless of memory availability.}

\subsection{Discussion}

The analysis of the impact of function calls monitoring on the user experience when the availability of the computational resources is limited revealed little influence of the computational resources. 
\begin{change} As a consequence, the logic of the monitoring can be activated and deactivated with limited attention to computational resources. Only in the case of CPU saturation higher than 90\%, monitoring should be avoided  since this could turn the application unresponsive.  \end{change}

Finally, results revealed that Instantaneous operations are less sensitive to memory availability compared to other kinds of operations. 


\begin{change}
\section{RQ3 - How do expert computer users react to the overhead produced by function calls monitoring, compared to the results obtained with RQ1 and RQ2?} 
\label{sec:empiricalstudy}
\end{change}

This research question investigates the coherence between the classification of the operations as resulting by the application of the criteria proposed by Seow and the feedback provided by actual users \begin{change}from our CS department\end{change}, on the applications and operations considered in our study. To this end, we asked a number of users to assess operations of different categories while exposed to a range of overhead values, and we compared the results to the ones obtained with the classification criteria by Seow. In the following, we present the design of the empirical study, the results, and their critical discussion. 


\subsection{Design}

We study how actual users perceive the system response time by considering operations exposed to overhead values in the ranges  $0-30\%$, $30-80\%$, $80-180\%$, $180+\%$ and operations belonging to the five operation categories used in our study (Instantaneous, Immediate, Continuous Simple, Continuous Complex, and Captive).
To expose every participant to the same interactions and to exactly the same overhead, we recorded videos showing the execution of the same operations executed for RQ1 and RQ2, while monitoring function calls. Each participant classifies each operation as either running slow or running in the expected amount of time.

%



The study involves 22 subjects who are members of our CS department, and include students, researchers, and professors. They all regularly use interactive applications. 
All the participants experienced all the operation categories and all the overhead values. 


\begin{table}[]
\centering
\scriptsize
\begin{tabular}{cccccc}
\toprule
\multirow{4}{*}{\textbf{Overhead range}}                         & \multicolumn{5}{c}{\textbf{Operation category}}                                                                                                                                                               \\  \cmidrule{2-6}
& \textbf{Instantaneous} & \textbf{Immediate} & \textbf{\begin{tabular}[c]{@{}l@{}}Continuous\\ Simple\end{tabular}} & \textbf{\begin{tabular}[c]{@{}l@{}}Continuous\\ Complex\end{tabular}} & \textbf{Captive} \\ \midrule
\textbf{0-30\%}         & 4                      & 2                  & 10                                                                   & 5                                                                     & 3                \\ \midrule
\textbf{30-80\%}        & 4                      &                    & 2                                                                    & 2                                                                     &                  \\  \midrule
\textbf{80-180\%}       & 1                      & 1                  & 3                                                                    &                                                                       &                  \\  \midrule
\textbf{180+\%}         & 1                      & 1                  & 1                                                                    &                                                                       &                  \\ \bottomrule
\end{tabular}
\caption{Number of operations for the different combinations of overhead ranges and categories.}
\label{table:op_distribution}
\vspace{-0.2cm}
\end{table}

For the evaluation, we selected eight meaningful tasks that cover the five possible operation categories and the four possible overhead ranges. Selecting fewer tasks would not allow us to cover enough cases, while using more cases would be too demanding for the participants. The distribution of the operations included in the tasks are shown in Table~\ref{table:op_distribution}.

The laboratory session started with the participants receiving an instruction sheet with general information about the structure of the experiment and a description of the tasks to be assessed, including  text and screen-shots to avoid misunderstandings. Subjects were not told about the specific aim of the experiment, nor about the presence of function calls monitoring.
After the video corresponding to a task has been reproduced and before moving to the next task, the participant evaluated the response time of each operation according to two possible levels: running slow or running as expected.

\subsection{Results}


\begin{figure}[h]
\centering
\includegraphics[width=\textwidth]{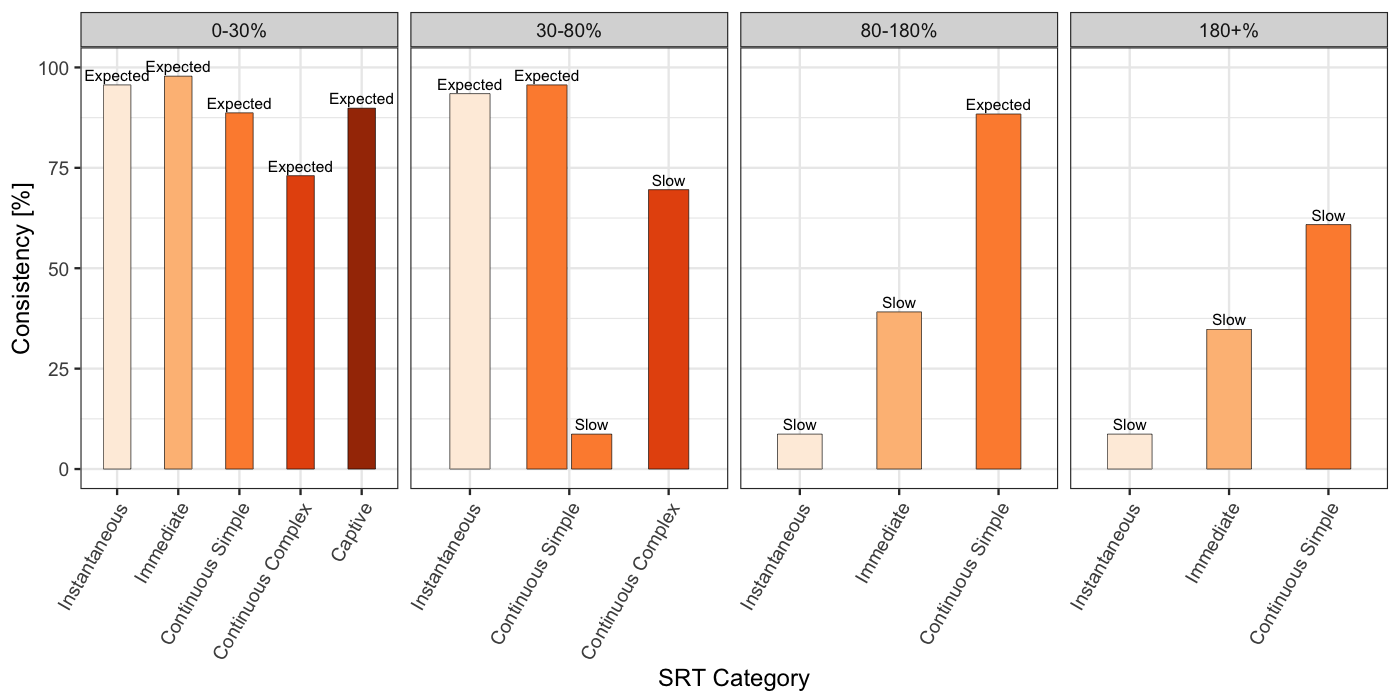}
\caption{Consistency in the evaluation of SRT between our approach and user perception.}
\vspace{-0.2cm}
\label{fig:new_slow_oh}
\end{figure}

To evaluate the consistency between the assessment based on the classification by Seow and the responses provided by the human subjects, we reclassified each operation assessed by users according to the classification by Seow and measure their coherence. 

Figure~\ref{fig:new_slow_oh} shows the results. Each bar is an operation category, the label at the top of the bar shows its classification as running slow or running as expected according to our definition of slow operation, and the percentage shows the number of participants who responded coherently with the label at the top. Based on these results, we conclude that:
\vspace{-0.1cm}
\begin{itemize}
\item \emph{Our classification strategy and the participants agree on the operations that should not be considered slow}. In fact, all the operations labeled as \emph{expected} are classified in the same way by a percentage of the participants ranging from 75\% to 100\%.
\item \emph{Our classification and the participants tend to agree on the continuous operations that should be considered slow}. In fact, there is agreement in considering Continuous Simple operations exposed to more than 80\% overhead and the Continuous Complex operations exposed to more than 30\% overhead as slow operations.
\item \emph{The participants tolerate overhead better than revealed by our classification for quick operations}. In fact, there is disagreement on the Instantaneous and Immediate operations exposed to overhead higher than 80\% 
\end{itemize}

\vspace{-0.3cm}
\subsection{Discussion}

We can conclude that identifying slow operations based on the Seow classification is conservative: the operations that we identify running as expected are fine also for the actual users; on the other hand, there might be operations that we consider too slow but are instead running as expected for the users. In practice, developers following the recommendations resulting from our work encounter into a negligible risk of introducing noticeable overheads in their applications.

%
%


\section{Threats to Validity} \label{sec:threats}

The main threat to {\it Internal Validity} of our empirical investigation is the usage of the system response time categories defined by Seow~\cite{seow} to identify the operations that have been slowed down up to a level that can be recognized by the users. Involving a significant number of human subjects in the evaluation and asking them to evaluate every individual operation that is executed, in different applications and contexts, is however nearly infeasible. This is why we decided to rely on a well-known categorization of user operations that let us work with a significant number of samples. 
To mitigate this issue, we investigated the coherence between our evaluation and the assessment performed by actual users for a subset of the operations and discovered that our findings provide a slightly conservative picture of how users perceive overhead. 

\begin{change}
Another potential threat is the representativeness of the participants we used in the empirical experiment to respond to RQ3. 
However, the use of popular applications also known to non-computer experts mitigates the potential bias introduced by the selected subjects.
\end{change}

Another potential threat is the choice of the individual operations that have been used in the study. Although we can potentially design the test cases in many different ways, we mitigated the issue of choosing the operations by focusing on the most relevant functionalities of each application, possibly including a large number of operations from most of the categories. 

The main threat to {\it External Validity} of our empirical evaluation is the generalizability of the results. The study focuses on function calls monitoring for regular desktop applications and, although some results might have a broader applicability, they should be interpreted mainly in that context. Considering other contexts require the replication of this study. 

\begin{change}
Another potential threat is related to the sample size of actual subjects we used in the empirical experiment to respond to RQ3. To further validate the results achieved, it is necessary to extend the experiment in such a way as to include many more subjects and potentially with different backgrounds. 
%
	
\end{change}


\section{Findings} \label{sec:findings}

In this section we summarize the main findings that result from the empirical experience reported in this paper:

\begin{itemize}
\item \textbf{Overhead up to 30\% is likely to be well tolerated by users}. Our results show that an overhead up to 30\% is seldom the cause of operations recognized as slow. 
This suggests that enriching applications running in the field with processes that collect data and analyze executions is feasible.

\item \textbf{Collecting sequences of function calls from the field is feasible in most of the cases}. Our results show that the actual overhead produced by function calls monitored is below 30\% in the vast majority of the cases. 
Furthermore, less than the 20\% of the executed operations are likely to be perceived as slowed down. Although the cases where the impact of monitoring is heavier must be carefully handled, results suggest that extensively collecting data about sequences of function calls from the field is possible.   

\item \textbf{Specific operations require special handling of monitoring features}. In our experiment we reported operations that presented an exceptionally high overhead. This happened both across categories and specifically in one application, that is, the Immediate operations present in Paint.NET. This result suggests that applications must be carefully analyzed before being instrumented so that the overhead introduced by the probes can be properly controlled, detecting these special cases.   

\item \textbf{Computational resources have little influence on the impact of monitoring}. Results show that CPU and RAM availability have not a significant impact on the relative cost of collecting function calls. It is however true that the cumulative effect of CPU load and monitoring may introduce a prohibitive overhead reaching a peak of more than 50\% of the operations perceived as slowed down, in contrast with the normal impact of monitoring that affected less than 20\% of the operations in the worst case.  

\item \textbf{Instantaneous operations are more resilient to overhead}. Instantaneous operations demonstrated to tolerate well overhead, also when the CPU is extremely busy. This is probably due to the intrinsic nature of these operations that can be executed fast almost without interruption, even if the CPU is busy. Moreover, human subjects demonstrated to tolerate particularly well the overhead introduced in Instantaneous operations.
\end{itemize}

These findings can be exploited by organizations that use experimentation techniques to improve their products and processes. Indeed, they can refine data collection  strategies to be less intrusive while collecting significant amount of data about the behavior of the software. 

In particular, our findings may impact:

\begin{itemize}
\item \emph{Requirements engineers}, who may exploit field monitoring solutions to profile users and evolve requirements following the usage scenarios discovered in the field. \begin{change}However, in order to be used without impacting on the user experience, the overhead introduced by profiling techniques should not exceed 30\%.\end{change} 

\item \emph{Software developers}, who may exploit highly-optimized monitoring procedures to collect software behavioral data, discovering how software is actually used in the field and improve their products, accordingly. \begin{change}Since computational resources have little influence on the impact of monitoring, environmental conditions for monitoring should not be a problem for developers.\end{change} 

\item \emph{Testers}, who may exploit fine-grained monitoring to collect accurate information about the behavior of a deployed product, with a specific focus on failures, to reveal and fix faults earlier. \begin{change} We demonstrated that function calls monitoring is feasible in most of the cases. If applied carefully it could be of great impact to the software testing community, considering that function calls monitoring has been widely used for several tasks, including debugging and fault failures reproduction~\cite{Murtaza:FaultLocalizationTheory:GTSE:2015,Jin:BugRedux:ICSE:2012}.\end{change} 

\item \emph{DevOps} architects, who engineer continuous monitoring solutions that can cost-effectively support continuous deployment, and the development process more in general\begin{change}, as long as they monitor between the boundaries we identified in this study, such as keeping overhead under 30\% and prioritizing Instantaneous operations because of their resilience to monitoring overhead.\end{change} 
\end{itemize}




\section{Related Work} \label{sec:related}

In this section we relate our work to software experimentation, to studies about the impact of SRT delays on the quality of the user experience, and to studies on field monitoring and analysis. 
%
\smallskip

Regarding \emph{software experimentation}, there are several approaches and studies that exploited software systems to collect actual evidence, especially using field data. A systematic way to collect field data is to perform randomised controlled experiments (e.g., A/B tests), for instance to study how a new feature or a change may impact the user experience. 
Continuous deployment is a well-known practice that benefits directly from controlled experiments~\cite{fabijan2017benefits}, for instance companies such as Microsoft reported to run more than 200 experiments concurrently every day within their products~\cite{kohavi2013online}. Relevantly, Kevic et al.~\cite{kevic2017characterizing} presented an empirical characterisation of an experimentation process when applied to the Bing web search engine, and Fagerholm et al.~\cite{fagerholm2017right} provided a model that enables continuous customer experiments aimed to software quality improvement. Our work relates to these studies because it provides evidence that can help engineers designing better data collection solutions that do not affect the user experience.

Regarding the \emph{perception of the SRT}, there are several studies in the context of the research in Human Computer Interaction (\textit{HCI}) and controlled experiments. 

The importance of controlled experiments has been extensively discussed and demonstrated. For instance, Fabijan et al.~\cite{fabijan2017benefits} reported the benefits of online controlled experiments for software development processes, studying how using customer and product data could support decisions throughout the product lifecyle.

Killeen et al.~\cite{killeen_optimal_1987} demonstrated that users are unlikely to recognize time variations inferior to 20\% of the original value. Our results are aligned with this study, since delays up to 30\% do not generate slowdowns recognizable by future users.

Ceaparu et al.~\cite{Ceaparu2004} studied how interactions with a personal computer may cause frustrations. In their experiment, users were asked to describe in a written form sources of frustration in human-computer interactions.
The participants declared that applications not responding in an appropriate amount of time and Web pages taking long time to process are the main sources of frustration in human-computer interactions, thus confirming the relevance of our investigation.

Other studies stressed user tolerance in specific settings.
For instance, Nah et al.~\cite{Nah2004} analyzed how long users are willing to wait for a Web page to be downloaded. The results of the experiment showed that users start noticing the slowdowns after two seconds delays and that do not tolerate slowdowns longer that 15 s.
A threshold of 15 s has been reported as the maximum that can be tolerated before perceiving an interruption in a conversation with an application also in other studies~\cite{Nielsen1999,Miller1968}.
An experiment conducted by Hoxmeier and Di Cesare~\cite{Hoxmeier00systemresponse} studied how fixed slowdowns (3, 6, 9, and 12 s) to interactions may affect the user appreciation and perception. Results show that a limit for the user tolerance is 12 s and that a linear relationship exists between SRT and user satisfaction. 

Kohavi et al.~\cite{kohavi2009controlled} show that slowdowns in Web applications may affect the user experience, causing loss of money for companies. For example, Amazon reported a loss of 1\% in sales because of a 100 ms slowdown, and Microsoft similarly, reported a loss of 1\% in user queries when adding a slowdown of one second to their search site.
Differently from these studies, our investigation does not aim to identify the maximum overhead that can be tolerated nor the cost to companies, but rather to identify delays and overhead levels that cannot be even recognized by users.


\smallskip

In the scope of \emph{monitoring techniques}, there are techniques that implemented mechanisms to limit the overhead introduced in the monitored system~\cite{cornejo2017flexible}. For instance, distributive monitoring can be used to divide the monitoring workload between several instances of a same application in order to lower the overhead introduced by monitoring activities~\cite{Orso:GammaSystem:ISSTA:2002,Bowring:MonitoringDeployedSoftware:PASTE:2002}. Briola et al.~\cite{Briola2014319,Mascardi2013300} and Ancona et al.~\cite{Ancona201635}  exploited a similar intuition to cost-efficiently monitor multi-agent systems.


Alternatively, information can be collected at run-time only with a given probability or according to a strategy~\cite{prepost}. This strategy has been exploited in the context of debugging~\cite{Jin:Sampling:SIGPLAN:2010,Liblit:BugIsolation:SIGPLAN:2003}, program verification~\cite{Bartocci:Adaptive:ICRV:2012}, and profiling~\cite{Hirzel:BurstyTracing:FDDO:2001}. Finally, monitoring can be optimized carefully balancing in-memory and saving operations~\cite{Cornejo:DelayedSaving:IEEEAccess:2019}.
The results reported in this paper can be exploited by these techniques and by practitioners~\cite{barik2016bones}, to further optimize their monitoring strategy, collecting more data without affecting users.

Depending on the kind of collected data, monitoring solutions may introduce overhead levels up to ~10000\%~\cite{Jin:BugRedux:ICSE:2012} as it is in the case of function calls monitoring. 
Also, slow software is one of the main reasons why users stop using applications, as reported in~\cite{Ceaparu2004,Hoxmeier00systemresponse,Miller1968}. Delaying too much some functionalities may cause loss of users and consequently the failure of the project.
The results obtained with our study may help practitioners to design context-aware techniques that achieve a better compromise between collecting data and impacting on the user experience, such as proposed in~\cite{Bartocci:Adaptive:ICRV:2012}. Our findings provide initial insights in this direction.


\section{Conclusions} \label{sec:conclusions}

Collecting data from the field is extremely useful to discover how applications are actually used and support software engineering tasks. For instance, several monitoring techniques collect sequences of function calls to reproduce failures~\cite{Jin:BugRedux:ICSE:2012}, detect malicious behaviors~\cite{Gorji:MalwareDetection:ACMSE:2014}, debug applications~\cite{Murtaza:FaultLocalizationTheory:GTSE:2015}, profile software~\cite{Elbaum:Profiling:TSE:2005}, optimize applications~\cite{Zhao:Optimization:OOPSLA:2014}, and mine models~\cite{Mariani:Revolution:ISSRE:2012,Mariani:DynamicAnalysis:TSE:2011}. If retrieving this data is indeed useful, knowing the impact of the monitoring activity on the user experience is also extremely important. In fact, monitoring techniques can be feasibly applied only if they work seamlessly.

This paper presented a study about the impact of function calls monitoring considering both the monitoring overhead and the operations that may be perceived as slowed down by the users. Results show that an overhead up to 30\% can be likely introduced in the operations without annoying users and that function calls monitoring often produce an overhead below this limit. We found however operations that are slowed significantly and that require special care when monitored. These findings suggest that monitoring capabilities cannot be introduced blindly, but they must be customized to the characteristics of the monitored program. Results also suggest that computational resources (RAM and CPU) have little influence on the impact of monitoring. 

Future work consists of exploiting the results obtained with this study to design monitoring techniques that can collect function calls from the field without being recognized by users. 
\begin{change}
We will further consider the categorization of the monitoring metrics we introduced in this work into useful usability groups, that can be reused at large scale experiments as recommended in the study by Rodden et al.~\cite{rodden2010measuring}.
\end{change}
Also, we will consider applications with longer response time, such as scientific experiments and job processing applications, studying the feasibility of monitoring in these fields.

\section*{Acknowledgment}
This work has been partially supported by the H2020 Learn project, which has been funded under the ERC Consolidator Grant 2014 program (ERC Grant Agreement n. 646867) and the GAUSS national research project, which has been funded by the MIUR under the PRIN 2015 program (Contract No. 2015KWREMX).

\bibliographystyle{elsarticle-num}
\bibliography{JSS}

\end{document}